\documentclass[11pt]{article}
\usepackage{amssymb,amsmath,amsfonts}
\usepackage{graphicx}
\usepackage{graphics}
\usepackage{eepic,epsfig}
\usepackage{xcolor}

\begin{document}
	\title{Thermal correction on the Casimir energy in a Ho{\v{r}}ava-Lifshitz Lorentz-violating scenario}
	\author{E. R. Bezerra de Mello$^{1}$%
		\thanks{%
			E-mail: emello@fisica.ufpb.br} ,\thinspace\ Herondy  Mota$^{1}$\thanks{E-Mail: hmota@gmail.com.br} \\
		\\
		$^{1}$\textit{Departamento de F\'{\i}sica, Universidade Federal da Para\'{\i}%
			ba}\\
		\textit{58.059-970, Caixa Postal 5.008, Jo\~{a}o Pessoa, PB, Brazil}}
	\maketitle
	
	\begin{abstract}
		The Casimir effect is considered a great triumph of Quantum Field Theory. Originally the Casimir energy was investigated considering the vacuum fluctuation associated with electromagnetic field; however it has also been analyzed considering different type of quantum fields. More recently the  Casimir energy was  studied in the context of Lorentz symmetry violation.  In this work we investigate the effect of finite temperature on the Casmir energy considering a Lorenttz violation symmetry in a Ho{\v{r}}ava-Lifshitz scenario.  In this sense we consider a massive scalar quantum field confined in a region between two large and parallel plates.  So our main objective is to investigate how, in this Lorentz violation scenario (LV), the Casimir energy depends on the temperature.  Another point to be analyzed, is influence of the parameter associated with the LV, $\xi$, in the thermal correction.  
	\end{abstract}

	\maketitle
	
	\section{Introduction}
	The Casimir effect is widely regarded as a triumph of Quantum Field Theory. It represents a macroscopic quantum force between two neutral conducting surfaces. The Casimir effect was theoretically proposed in a pioneer paper in $1948$ by H. B. Casimir in \cite{Casimir:1948dh}.  Originally this effect was analyzed considering the vacuum state associated with the electromagnetic field confined between two large, parallel, conducting and neutral plates; however, it has been extended considering different type of quantum fields. More recently the Casimir energy has been studied in the context of Lorentz symmetry violation (LV). In this sense two different approaches have been taking into consideration. The $CPT-$even, aether-like approach is based on a spontaneous violation of the Lorentz symmetry, that is implemented through the emergence of expected values of nonzero vacuum by some vector and tensor components, which implies in preferential directions, therefore, space-time anisotropy. The other formalism, named  Ho{\v{r}}ava-Lifshitz (HL) one \cite{hovrava2009quantum}, was proposed as an attempt to construct a renormalizable quantum field theory for gravity. In this formalism, the propagator for  gravitons  depends on the energy scale, introducing an anisotropy between space and time coordinates. In addition, D. Anselmi  in \cite{anselmi2009weighted} and \cite{anselmi2009weighted2}, proposed a model  that violates Lorentz symmetry explicitly at high energies and is renormalizable by weighted power counting. The model contains higher space derivatives, which improve the behavior of propagators at large momenta, but no higher time derivatives.

	The analysis of the influence of LV in a aether-like scenario in the Casimir effect considering massive scalar quantum field  has been developed in \cite{Cruz:2017kfo}. There it was considered that the field obeys Dirichlet, Neumann and mixed boundary conditions on the two parallel plates, and that the breaking of Lorentz symmetry is implemented by the presence of a time-like and/or space-like unity vector that couples with the derivative of the scalar field. Moreover, the analysis of the influence of temperature in the Casimir effect in this context has been investigated in  \cite{Cruz:2018bqt}.  In \cite{Erdas2020} the author analyzed the influence of a constant magnetic field parallel to the plates, on the Casimir effect in this $CPT-$even violation of Lorentz symmetry, considering charged scalar field.

	The studies of Casimir effects associated with massless scalar quantum field confined in the region between two large parallel plates taken into account the HL formalism, has been developed in \cite{Ferrari,Ulion}; in addition the Casimir effect associated with a massive real scalar field was analyzed  in \cite{Maluf2020}. Recently the Casimir effects associated with a massive self-interacting scalar quantum field in a HL Lorentz-violation scenario, considering the Dirichlet boundary condition on the two large and parallel plates,  has been investigated in \cite{Farias}.  Finally the analysis of Casimir energy in a new LV scenario, corresponding  to the presence of an arbitrary constant space-like vector in a CPT-even aether-like approach, considering a direct coupling between this vector with the derivative of the field in higher order has been developed recently in \cite{Robson}.
	
	In this work we want to investigate the effect of finite temperature on the Casimir energy considering a Lorentz violation symmetry in a Ho{\v{r}}ava-Lifshitz scenario. In this way, we consider a massive scalar quantum field with a nonzero chemical potential, $\mu$, in thermodynamic equilibrium with a thermal bath possessing  a finite temperature, $T=\beta^{-1}$. In order to develop this analysis we assume that the quantum field is confined in a region between two large and parallel plates, by imposing on this field the Dirichlet, Neumann and mixed boundary conditions on the plates. So our main objective is to investigate how, in this LV scenario, the Casimir energy depends on the temperature.

	This paper is organized as follows. In Section \ref{the_model}, we present the modified Klein-Gordon Lagrangian within the framework of the Ho{\v{r}}ava-Lifshitz scenario. In Section \ref{Thermal_correction}, we calculate the thermal corrections associated with the Casimir effects. Considering initially that the scalar field obeys Dirichlet boundary condition on two large and parallel plates in subsection \ref{Dirichlet}, we  obtain an expression for the dispersion relation in terms of the discretized momentum perpendicular to the plates. Assuming that the field is in thermal equilibrium at finite temperature $T$, we will see that the ensemble average of the Hamiltonian operator is expressed in terms of its vacuum expectation values and a finite temperature contribution coming from the particle and anti-particle excitation. Because the vacuum expectation value of the Hamiltonian has been analyzed before for massless \cite{Ferrari,Ulion} and massive \cite{Maluf2020} quantum scalar field, here we are mainly interested to calculate the thermal contribution. By using the Abel-Plana summation formula to take into account the sum over the discretized momentum, we present the thermal correction as the  summation of three distinct contributions. These contributions are analyzed separately. Our most important objective is to calculate the thermal correction of the Casimir energy.
	Unfortunately this quantity is presented by an unsolvable integral representation. Because this representation is nor very enlightening, we present graphs exhibiting its behavior for different values of the critical exponent, $\xi$, associated with the violation of Lorentz symmetry in the HL context, and  also considering different values of temperature.  In subsection \ref{Neumann}, we consider that the quantum field obeys Neumann boundary condition on the two large and parallel plates. As we will see, this case is very similar to the previous one, so we briefly discuss about the thermal correction to the Casimir energy. Finally in subsection \ref{Mixxed}, we consider that the field obeys a mixed boundary condition. For this case, the thermal correction to the Casimir energy presents small but important modifications when compared with the two previous analysis. Also, for the thermal correction of the Casimir energy our result is expressed in terms of not solvable integral, and only numerical analysis can provide the behavior of this observable for different values of temperature and critical exponent.  The most relevant results are summarized and discussed in section \ref{Concl}. In the Appendices \ref{Ap_a} and \ref{Ap_b} we provide the steps needed to express thermal energy correction in a closed form for the case where the Lorentz symmetry is preserved.  Here, we adopt natural units $\hbar = c =k_B= 1$, and the metric signature convention $(+,-,-,-)$.

	\section{The model setup}
	\label{the_model}
	
	In this section, we will introduce the theoretical model in which the Klein-Gordon equation is modified in the context of the Ho{\v{r}}ava-Lifshitz (HL) formalism. Our main objective is to obtain the expression for the field operator, that satisfies Dirichlet boundary condition on two parallel plates with areas $L^2$, separated by a distance $a$, being $a \ll L$. The action associated with this system reads, 
	\begin{equation}
		\label{S}
		S = \frac{1}{2} \int \mathrm{d}t \mathrm{d}^d x \left( \partial_{0}\phi\partial_{0}\phi
		- l^{2(\xi-1)}\partial_{i_{1}}\partial_{i_{2}}...\partial_{i_{\xi}}
		\phi\partial_{i_{1}}\partial_{i_{2}}...\partial_{i_{\xi}}\phi-m^2\phi^2\right) \  ,
	\end{equation}
	where $l$ is a parameter with dimension of length.\footnote{In fact the parameter $l$ is of the order of the inverse of the energy scale where the Lorentz symmetry is broken.} The space-time anisotropy caused by the breaking of Lorentz symmetry, 	occurs due to different properties of scales in which coordinates space and time are set, so that 	the theory is invariant under the rescaling $x\to bx$, $t\to b^{\xi}t$, where $\xi$ is a number 
	called the critical exponent \cite{hovrava2009quantum}.  In fact the Lorentz invariance is broken for $\xi\neq 1$; however the main objective for this violation is that for a convenient choice of $\xi$ bigger than unity, the theory becomes free of ultraviolet divergence. To avoid problem related with fractional derivative in this work we will assume $\xi$ being an integer number; however our results are analytic functions of $\xi$ and no restrictions are needed.
	
	In the case $(3 + 1)-$dimensions the modified Klein-Gordon equation reads,
	\begin{equation}
		\label{kgmodific}
		[\partial_{0}^2 + l^{2(\xi -1)}(-1)^{\xi}({\vec{\nabla}^2})^\xi +m^2]\phi = 0  \   .
	\end{equation}
	
		Here we will consider an hermitian quantum field confined in a region between two parallel flat boundaries. Moreover, we will assume that the whole system is in thermal equilibrium at temperature $T = 1/ \beta$. As we know finite temperature changes the properties of the quantum vacuum state, introducing thermal fluctuations and excitations, consequently, there is no pure state but a statistical mixture represented by the density operator
	\begin{eqnarray}
		\label{density_operator}
		\hat{\rho} = \frac{e^{-\beta(\hat{H} - \mu \hat{N})}}{Z} \  ,
	\end{eqnarray}
	where the grand canonical partition function, $Z$, is defined by
	\begin{eqnarray}
		\label{grand_partition_funtion}
		Z = Tr \Big{[}e^{-\beta(\hat{H} - \mu \hat{N})}\Big{]} \  .
	\end{eqnarray}
	In both expressions above, $\hat{H}$ represents the Hamiltonian operator and $\hat{N}$ the number operator. The parameter $\mu$ is the chemical potential.  In fact for system containing real particles in thermal equilibrium, the chemical potential corresponds to a changing of free energy of a thermodynamical process when particles are added or removed.
	
	The ensemble average of any physical quantity represented by an operator $\hat{O}$ can be computed by application of the formula
	\begin{eqnarray}
		\langle \hat{O} \rangle = {\rm Tr} \left(\hat{\rho} \hat{O}\right).
	\end{eqnarray}
	For the product of creation and annihilation operators we have,
	\begin{eqnarray}
		\label{average_1}
		Tr\left(\hat{\rho}\hat{a}^{\dagger}_\sigma \hat{a}_{\sigma'} \right)= \frac{\delta_{\sigma,\sigma'}}{e^{\beta(\omega_n - \mu)}-1} \  ,
	\end{eqnarray}
	where $\omega_n$ represents single particle levels energy. For bosonic field we must have $\omega_0 \geq \mu$, being $\omega_0$ the minimum of energy. 
	
For the ensemble average of the Hamiltonian considered we have,
	\begin{eqnarray}
		\label{H-Thermal}
		E=\langle \hat{H} \rangle=E_0+E_{T} \ , 
	\end{eqnarray}
	where $E_0$ represents the vacuum energy and $E_T$, the thermal correction. The vacuum energy associated with this configuration has been analyzed in \cite{Maluf2020}. Our objective here is to analyze the thermal contribution. 
	
\section{Thermal correction for the Casimir effect}
	\label{Thermal_correction} 
	
In this section, we will investigate the influence of the temperature on the Casimir energy in a HL Lorentz violation scenario. In this way we will consider that the field is confined in a region between two large and parallel plates, admitting that, on these surface, the field obeys Dirichlet, Neumann and mixed boundary conditions.
	
\subsection{Dirichlet boundary condition}
\label{Dirichlet}	
	Admitting that the field operator obeys Dirichlet boundary condition on the parallel plates, placed perpendicular to the $z-$axis at points $z=0$ and $z=a$, i.e., 
	\begin{eqnarray}
		\label{Dirichlet_condition}
		\phi(x)\Big{|}_{z=0} = \phi(x)\Big{|}_{z=a}=0 \  , 
	\end{eqnarray}
	and adopting the standard procedure described in textbooks on Quantum Field Theory \cite{Mandl}, one finds the field operator:
	\begin{eqnarray}
		\label{Dirichlet_operador_campo}
		\hat{\phi}(x) = \int d^2\vec{k} \sum_{n=1}^{\infty} { \frac{1}{\sqrt{(2\pi)^2ak_0}} }
		\sin \Big{(} \frac{n\pi}{a}z \Big{)} \Big{[} \hat{a}_n(\vec{k})e^{-ikx}+\hat{a}^{\dagger}_n(\vec{k})e^{ikx} \Big{]},
	\end{eqnarray}
	where
	\begin{eqnarray}
		\label{k_def}
		kx=k_0x_0-k_{x}x-k_{y}y \  ,
	\end{eqnarray}
	with 
	\begin{eqnarray}
		\label{k0_def}
		k_0=\omega_{{\vec{k}},n}=l^{\xi-1}\sqrt{(k_x^2+k_y^2+(n\pi/a)^2)^\xi+v^{2\xi}} \  ,  
	\end{eqnarray}
	being ${\cal}{v}^\xi=ml^{1-\xi}$.
	
	In (\ref{Dirichlet_operador_campo}),  $\hat{a}_{n}(\vec{k})$ and $\hat{a}^{\dagger}_{n}(\vec{k})$ represent the annihilation and creation operators, respectively,  characterized by the set of quantum numbers $\sigma = \{k_{x},k_{y},n\}$.  In addition, the Hamiltonian operator, $\hat{H}$, for this system reads,
	\begin{equation}
		\label{Hamilt_0}
		\hat{H} = \frac{1}{2}\int \mathrm{d}^2\textbf{k} \sum_{n=1}^{\infty} 
		\omega_{\textbf{k},n}\left[2 a^{\dagger}_{\textbf{k,n}} a_{\textbf{k,n}} + 
		\frac{L^2}{(2\pi)^2}\right] \   .
	\end{equation}

	By using  the relation \eqref{average_1}, we have:
	\begin{eqnarray}
		\label{energy}
		E_{T}=\frac{L^2}{(2\pi)^2}\int{d^{2}\vec{k}}\sum_{n=1}^{\infty}\frac{\omega_{\vec{k},n}}{e^{\beta(\omega_{\vec{k},n}-\mu)}-1} \  .
	\end{eqnarray}
	Using the expression
	\begin{eqnarray}
		\label{identidade}
		(e^{z}-1)^{-1}=\sum_{j=1}^{\infty}e^{-jz} \  ,
	\end{eqnarray}
	the equation \eqref{energy}, can be rewritten as,
	\begin{eqnarray}
		\label{energyT}
		E_T=\frac{L^2}{(2\pi)^2}\sum_{j=1}^{\infty}e^{j\beta \mu}\int_{0}^{\infty}{d^2k}\sum_{n=1}^{\infty}\omega_{\vec{k},n}e^{-j\beta \omega_{\vec{k},n}}  \   .
	\end{eqnarray}
	
	In order to develop the summation on the quantum number $n$, we will use the Abel-Plana summation formula \cite{{Bordag:2009zzd,Saharian:2007ph}}:
	\begin{eqnarray}
		\label{AP1}
		\sum_{n=1}^{\infty}F(n)=-\frac{1}{2}F(0)+\int_{0}^{\infty}{dt}F(t)+i\int_{0}^{\infty}\frac{dt}{e^{2\pi t}-1}\left[F(it)-
		F(-it)\right]  \  .
	\end{eqnarray}
	For our case,
	\begin{eqnarray}
		F(n):=\omega_{\vec{k},n} e^{-j\beta \omega_{\vec{k},n}} \  , \ \omega_{{\vec{k}},n}=l^{\xi-1}\sqrt{(k_x^2+k_y^2+(n\pi/a)^2)^\xi+v^{2\xi}} \ .
	\end{eqnarray}
	
	In what follows we will analyze each contribution to the thermal energy contained in the development presented by the sum. 
	
	\subsubsection{First contribution}
	\label{Cont_1}
	The first term on the r.h.s of \eqref{AP1} provides the contribution to $E_T$ due to a single plate. After some intermediate steps, it reads,
	\begin{eqnarray}
		\label{E_T_1_a}
		E_T^{(1)}=-\frac{L^2l^{\xi-1}}{4\pi}\sum_{j=1}^{\infty}e^{j\beta \mu}\int_{0}^{\infty}dk k \sqrt{k^{2\xi}+v^{2\xi}} \ e^{-j\beta l^{\xi-1}\sqrt{k^{2\xi+v^{2\xi}}}}   \ .
	\end{eqnarray}
	
	Defining a new variable $x=k^\xi$, the integral over the variable $k$ above  can be rewritten as shown below:
	\begin{eqnarray}
		\label{Int_1}
		{\cal{I}}=\frac1\xi\int_0^\infty dx x^{2/\xi-1}\sqrt{x^2+\sigma^2} \ e^{-\alpha\sqrt{x^2+\sigma^2}} \  , 
	\end{eqnarray}
	where, using a short notation,  we have defined $\sigma=v^\xi$  and $\alpha=j\beta l^{\xi-1}$. Using  the integral representation below for the Macdonald function \cite{Abra}: 
	\begin{eqnarray}
		\label{Macd_fun}
		K_\nu(xz)=\frac{\sqrt{\pi}}{\Gamma(\nu+1/2)}\left(\frac{x}{2z}\right)^\nu\int_0^\infty dt \ t^{2\nu} \frac{e^{-x\sqrt{t^2+z^2}}}{\sqrt{t^2+z^2}}  \  ,
	\end{eqnarray}
	we can see that 
	\begin{eqnarray}
		\label{Ident_1}
		{\cal{I}}=\frac1\xi\frac{\partial^2}{\partial\alpha^2}\int_0^\infty dx x^{2/\xi-1}\frac{e^{-\alpha\sqrt{x^2+\sigma^2}}}{\sqrt{x^2+\sigma^2}} \  .
	\end{eqnarray}
	Nonetheless, we can express the integral above in terms of the Macdonald function assuming $\nu=\frac1\xi-\frac12$. After some intermediate steps, we get:
	\begin{eqnarray}
		\label{Ener_1}
		E_T^{(1)}&=&-\frac{L^2l^{\xi-1}}{4\pi\xi}\frac{2^{1/\xi}\Gamma(1/\xi)}{\sqrt{2\pi}}\sum_{j=1}^\infty e^{j\beta\mu}\left[v^{1+3\xi/2}\frac{K_{1/\xi+3/2}(j\beta m)}{(j\beta l^{\xi-1})^{1/\xi-1/2}}\right.\nonumber\\
		&-&\left.v^{1+\xi/2}\frac{K_{1/\xi+1/2}(j\beta m)}{(j\beta l^{\xi-1})^{1/\xi+1/2}}\right]  \  .
	\end{eqnarray}
	
	From the above result we can write its limit for:\\
	$i)$ Massless fields, $m=0$. In this case  we have to take $\mu=0$.  With the help of \cite{Grad}, we obtain,
	\begin{eqnarray}
		\label{E_T_1_b}
		E_T^{(1)}=-\frac{L^2l^{\xi-1}}{4\pi}\frac{\Gamma\left(1+2/\xi\right)}{\xi} \frac1{(\beta l^{\xi-1})^{1+2/\xi}}\sum_{j=1}^{\infty}\frac1{j^{1+2/\xi}} \  ,
	\end{eqnarray}
	where $\Gamma(z)$ represents the Gamma function \cite{Abra}. The summation over $j$ can be represented by the Riemann zeta function, $\zeta(z)$. Our final result is:
	\begin{eqnarray}
		\label{E_T_1_c}	
		E_T^{(1)}=-\frac{L^2l^{2(1/\xi-1)}}{4\pi}\frac{\Gamma\left(1+2/\xi\right)}{\xi}\frac1{\beta^{1+2/\xi}}\zeta\left(1+2/\xi\right) \  , \  \beta=1/T \ .
	\end{eqnarray}
	$ii)$ Case with $\xi=1$, i.e., Lorentz symmetry preserved. Using formulas from \cite{Grad}, the expression reads,
	\begin{eqnarray}
		\label{E_xi=1}
		E_T^{(1)}=-\frac{L^2}{4\pi}\sum_{j=1}^{\infty}e^{-j\beta(m-\mu)}\left(\frac{m^2}{j\beta}+ \frac{2m}{(j\beta)^2}+\frac2{(j\beta)^3}\right) \  .
	\end{eqnarray}
	The above summation can be represented in terms of special functions given in \cite{Abra}:
	\begin{eqnarray}
		E_T^{(1)}&=&-\frac{L^2}{4\pi}\left\{-\frac{m^2}{\beta}\ln\left(1-e^{-\beta(m-\mu)}\right)+\frac{2m}{\beta^2}
		{\rm polylog}\left(2,e^{-\beta(m-\mu)}\right)\right.\nonumber\\
		&+&\left.\frac{2}{\beta^3}{\rm polylog}\left(3,e^{-\beta(m-\mu)}\right)\right\}	\  \  .
	\end{eqnarray}
	
	\subsubsection{Second contribution}
	\label{Cont_2}
	Substituting the second term of \eqref{AP1} into \eqref{energyT}, we will obtain the contribution to $E_T$ in the absence of plates. It reads,
	\begin{eqnarray}
		\label{E_T_2_a}
		E_T^{(2)}=\frac{L^2}{(2\pi)^2}\sum_{j=1}^\infty e^{j\beta\mu}\int d^2k\int_0^\infty dt F(t) \  , 
	\end{eqnarray}
	with
	\begin{eqnarray}
		F(t)=l^{\xi-1}\sqrt{(k^2+(t\pi/a)^2)^\xi+v^{2\xi} }\  e^{-j\beta l^{\xi-1}\sqrt{(k^2+(t\pi/a)^2)^\xi+v^{2\xi}}} \  .
	\end{eqnarray}
	Defining a new variable $u=t\frac{\pi}{a}$, integrating over the angular variable in the two-dimensional $(k_x,k_y)$ plane, we get, 
	\begin{eqnarray}
		\label{E_T_2_b}
		E_T^{(2)}=\frac{L^2l^{\xi-1}a}{2\pi^2}\sum_{j=1}^\infty e^{j\beta\mu}\int_0^\infty dk k \int_0^\infty du \sqrt{(k^2+u^2)^\xi+v^{2\xi}} \ e^{-j\beta l^{\xi-1}\sqrt{(k^2+u^2)^\xi+v^{2\xi}}}   \  .
	\end{eqnarray}
	Changing the coordinate on the plane $(k,u)$ to polar one $(\sigma,\theta)$, after the integration over the angular variable, we get,
	\begin{eqnarray}
		\label{E_T_2_c}
		E_T^{(2)}=\frac{L^2l^{\xi-1}a}{2\pi^2}\sum_{j=1}^\infty e^{j\beta\mu}\int_0^\infty d\sigma \sigma^2\sqrt{\sigma^{2\xi}+v^{2\xi}} \ e^{-j\beta l^{\xi-1}\sqrt{\sigma^{2\xi}+v^{2\xi}}}  \  .	
	\end{eqnarray}
	
	The integral over the variable $\sigma$ can be expressed in terms of Macdonald function, similarly what we have done in previous subsection, defining a new variable $x$ by $\sigma^\xi=x$. Using the integral representation \eqref{Macd_fun}, and adopting $\nu=3/(2\xi)-1/2$, we obtain that the energy density, $\cal{E}$, reads,
	\begin{eqnarray}
		\label{Ener_2}
		{\cal{E}}=\frac{E_T^{(2)}}{aL^2}&=&\frac{l^{\xi-1}}{2\pi^2\xi}\frac{2^{3/2\xi}\Gamma(3/2\xi)}{\sqrt{2\pi}}\sum_{j=1}^\infty e^{j\beta\mu}\left[v^{3(1+\xi)/2}\frac{K_{3/2\xi+3/2}(j\beta m)}{(j\beta l^{\xi-1})^{3/2\xi-1/2}}\right.\nonumber\\
		&-&\left.v^{(3+\xi)/2}\frac{K_{3/2\xi+1/2}(j\beta m)}{(j\beta l^{\xi-1})^{3/2\xi+1/2}}\right]  \  .
	\end{eqnarray}
	
	Here we also can obtain the massless limiting case ($\mu=0$). Using \cite{Grad},  the energy density becomes,
	\begin{eqnarray}
		\label{E_T_2_d}
		{\cal{E}}=\frac{E_T^{(2)}}{aL^2}=\frac{l^{3/\xi-3}}{2\pi^2}\frac{\Gamma\left(1+3/\xi\right)}{\xi}\frac1{\beta^{1+3/\xi}} \zeta\left(1+3/\xi\right) \  .
	\end{eqnarray}
	Taking $\xi=1$ in \eqref{E_T_2_d}, we have,
	\begin{eqnarray}
		\label{Energy_density}
		{\cal{E}}=\frac	1{30}\,{\frac {{\pi }^{2}}{{\beta}^{4}}}  \  .
	\end{eqnarray}

	\subsubsection{The Casimir energy thermal correction}
	\label{Casimir}
	In this subsection we will calculate the Casimir thermal correction. This energy is obtained by substituting the third term of \eqref{AP1} into \eqref{energyT}. Adopting this procedure, we obtain:
	\begin{eqnarray}
		\label{Cas_1}
		E_{T,C}&=&\frac{iL^2l^{\xi-1}}{2\pi^2}\frac a\pi\sum_{j=1}^\infty e^{j\beta\mu}\int d^2{\vec{k}}\int_0^\infty \frac {du}{e^{2au}-1} \left[\sqrt{\left(k^2+(iu)^2\right)^\xi+v^{2\xi}}e^{-j\beta l^{\xi-1}\sqrt{\left(k^2+(iu)^2\right)^\xi+v^{2\xi}}}\right.\nonumber\\
		&-&\left.\sqrt{\left(k^2+(-iu)^2\right)^\xi+v^{2\xi}}e^{-j\beta l^{\xi-1}\sqrt{\left(k^2+(-iu)^2\right)^\xi+v^{2\xi}}}\right]  \  ,
	\end{eqnarray}
	where we have introduced a new variable $u=\pi t/a$. In order to develop the integration over the variable $u$, we will use the identity below,
			\begin{eqnarray}
				\label{identity1}
			\sqrt{\left(k^2+(e^{\pm i\pi/2}u)^2\right)^\xi+v^{2\xi}}=\left\{ 
			\begin{array}{ll}
				\sqrt{\left(k^2-u^2\right)^\xi+v^{2\xi}} \ , \  \  \  \   \  \ {\rm for} \  u<\sqrt{k^2+v^2} , & \\
			e^{\pm i\pi\xi/2}\sqrt{(u^2-k^2)^{2\xi}-v^{2\xi}} , \ {\rm for} \  u>\sqrt{k^2+v^2}  \ .
		\end{array}
		\right.
		\end{eqnarray}
		Dividing the integral with respect to this variable into two sub-intervals: from $[0, \sqrt{k^2+v^2}]$ and from $[\sqrt{k^2+v^2}, \infty)$.  We can see that the first integral vanishing remaining only the contribution in the second sub-segment; moreover, integrating over the angular variable on the plane $(k,\varphi)$, and developing some intermediate steps, we obtain,
	\begin{eqnarray}
		\label{Cas_2}
		E_{T,C}&=&-\frac{aL^2l^{\xi-1}}{\pi^2}\sin(\pi\xi/2)\sum_{j=1}^\infty e^{j\beta\mu}\int_0^\infty dk k\int_{\sqrt{k^2+v^2}}^\infty \frac {du}{e^{2au}-1} \sqrt{\left(u^2-k^2\right)^\xi-v^{2\xi}}\nonumber\\
		&\times&\cos\left(j\beta l^{\xi-1}\sqrt{\left(u^2-k^2\right)^\xi-v^{2\xi}}\right)  \  .
	\end{eqnarray}
	Defining a new variable $\rho=\sqrt{k^2+v^2}$, then we have $kdk=\rho d\rho$, so the above integral can be written as:
	\begin{eqnarray}
		\label{Cas_3}
		E_{T,C}&=&-\frac{aL^2l^{\xi-1}}{\pi^2}\sin(\pi\xi/2)\sum_{j=1}^\infty e^{j\beta\mu}\int_v^\infty d\rho \rho\int_{\rho}^\infty \frac {du}{e^{2au}-1} \sqrt{\left(u^2-\rho^2+v^2\right)^\xi-v^{2\xi}}\nonumber\\
		&\times&\cos\left(j\beta l^{\xi-1}\sqrt{\left(u^2-\rho^2-v^2\right)^\xi-v^{2\xi}}\right)  \  .
	\end{eqnarray}
	Defining $u=\rho y$, we get,
	\begin{eqnarray}
		\label{Cas_4}
		E_{T,C}&=&-\frac{aL^2l^{\xi-1}}{\pi^2}\sin(\pi\xi/2)\sum_{j=1}^\infty e^{j\beta\mu}\int_v^\infty d\rho \rho^2\int_1^\infty \frac {dy}{e^{2a\rho y}-1} \sqrt{[\rho^2(y^2-1)+v^2]^\xi-v^{2\xi}}\nonumber\\
		&\times&\cos\left(j\beta l^{\xi-1}\sqrt{[\rho^2(y^2-1)+v^2]^\xi-v^{2\xi}}\right)  \  .
	\end{eqnarray}
Finally defining a new variable $\rho=vz$, we have:
	\begin{eqnarray}
	\label{Cas_5}
	E_{T,C}&=&-\frac{L^2(am)^{1+3/\xi}\rho_\xi^3}{a^3\pi^2}\sin(\pi\xi/2)\sum_{j=1}^\infty e^{j\beta\mu}\int_1^\infty dz z^2\int_1^\infty \frac {dy}{e^{2(ma)^{1/\xi}\rho_\xi z y}-1} \nonumber\\
	&\times&\sqrt{[z^2(y^2-1)+1]^\xi-1}\cos\left(j\beta m\sqrt{[z^2(y^2-1)+1]^\xi-1}\right)  \ ,
\end{eqnarray}
being $\rho_\xi=(a/l)^{1-1/\xi}$. 

 At this point we would like to make a few comments about the role of the critical exponent on the thermal Casimir energy. The first one concerns to the presence of the the sine function in front of Eq. \eqref{Cas_5}. Although this function also appears in the zero-temperature Casimir energy, it is consequence of the identity \eqref{identity1}. The changing in the sign of the thermal Casimir energy for odd values of $\xi$ can be physically explained due to different dynamics obeyed by the scalar field, furnishing  different dispersions relations. Another peculiarity is that for even values of $\xi$ there is no interaction between the two large plates, on the other hand for odd values of the critical exponent the interaction may be attractive or repulsive. A second point that deserves to be mentioned  is related with the intensity of the thermal Casimir energy. Due to the dependence of \eqref{Cas_5} with the parameter $\rho_\xi$, that appears in the exponent of the denominator of the integrand over the variable $y$, we see that this denominator increases with $\xi$ once that the ration $a/l$ is much larger than unity for a realistic scenario, reducing significantly the intensity of the energy for higher values of $\xi$.

In Appendix \ref{Ap_a} we present the procedure adopted to obtain a closed expression for \eqref{Cas_5} considering $\xi=1$. The final expression is:
\begin{eqnarray}
	\label{Ener-Closed}
E_{C,T}&=&-\frac{L^2am^4}{\pi^2}\sum_{j=1}^\infty e^{j\beta \mu}\frac{\partial}{\partial\alpha}\left[\alpha\sum_{n=1}^\infty \frac{K_2\left(\sqrt{4n^2(ma)^2+\alpha^2}\right)}{(4n^2(ma)^2+\alpha^2)}\right]|_{\alpha=j\beta m} \  . 
\end{eqnarray}

 For massless field \eqref{Cas_4} takes a much simpler form:
	\begin{eqnarray}
		\label{Cas_6}
		E_{T,C}&=&-\frac{L^2\delta_\xi}{a^3\pi^2}\sin(\pi\xi/2)\sum_{j=1}^\infty \int_0^\infty dz z^{2+\xi}\int_1^\infty \frac {dy}{e^{2zy}-1} (y^2-1)^{\xi/2}\nonumber\\
		&\times&\cos\left(j\beta\delta_\xi z^\xi(y^2-1)^{\xi/2}/a \right)  \  , \ \delta_\xi=(l/a)^{\xi-1} \  ,
	\end{eqnarray}
	where we have defined $z=a\rho$.
	
In \cite{Cruz:2018bqt}, it is given a closed result for \eqref{Cas_6} considering $\xi=1$. For the sake of completeness, below we provide this expression:
 \begin{eqnarray}
	\label{Cas_6a}
	{E_{T,C}}& =& - \frac{L^2}{8 \pi ^2 a^3} \sum_{j=1}^{\infty} \frac{1}{\alpha_j^4}  \Bigg{[} 24 - 
	4\pi \alpha_j \text{coth} \Big{(}\frac{\pi \alpha_j}{2}\Big{)} - 2\pi^2\alpha_j^2\text{csch}^2\Big{(}\frac{\pi \alpha_j}{2}
	\Big{)} \nonumber \\ 
	& -& \pi^3\alpha_j^3\text{coth}\Big{(}\frac{\pi \alpha_j}{2}\Big{)}\text{csch}^2\Big{(}\frac{\pi \alpha_j}{2}
	\Big{)} \Bigg{]}  \  ,
\end{eqnarray}
being $\alpha_j=j\beta/a$. Unfortunately there is no closed expression for the summation over $j$. \footnote{In fact in \cite{Cruz:2018bqt} it is provide the thermal Casimir energy in a aether-like LV scenario, considering Dirichlet, Newman and mixed boundary conditions. The expression presented in \eqref{Cas_6a}, corresponds to the Dirichlet boundary condition case neglecting the parameter associated with the LV.}

\begin{figure}[!htb]
	\begin{center}
		\includegraphics[scale=0.35]{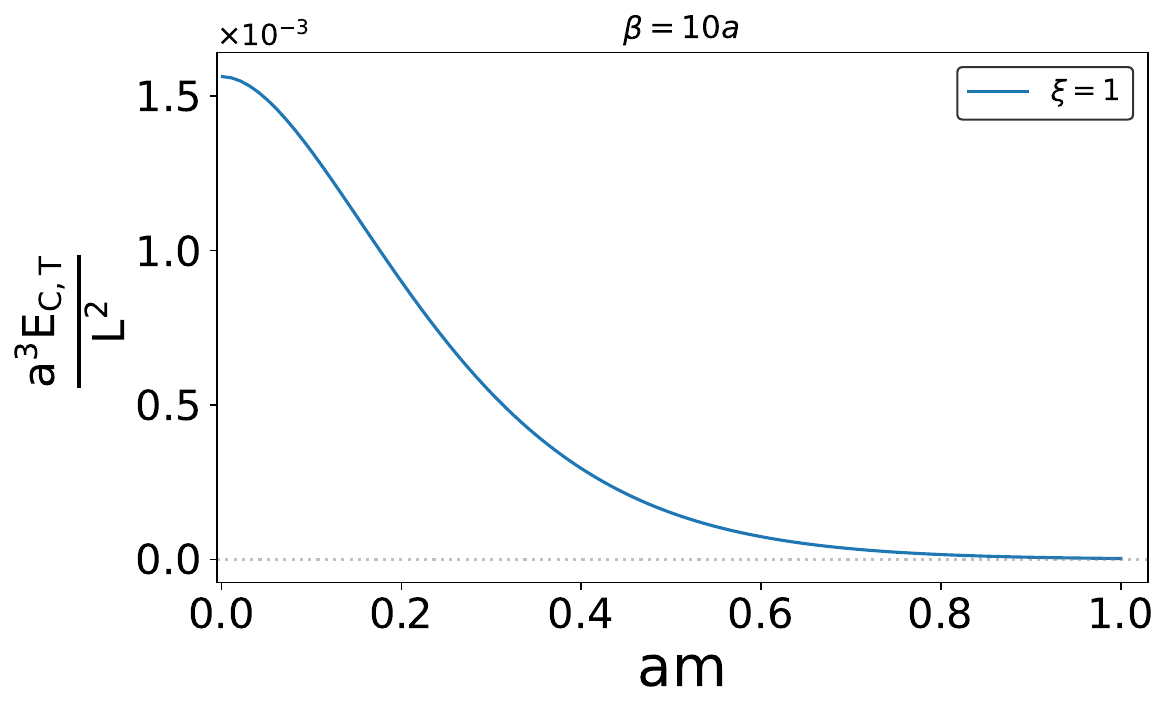}
		\quad
		\includegraphics[scale=0.35]{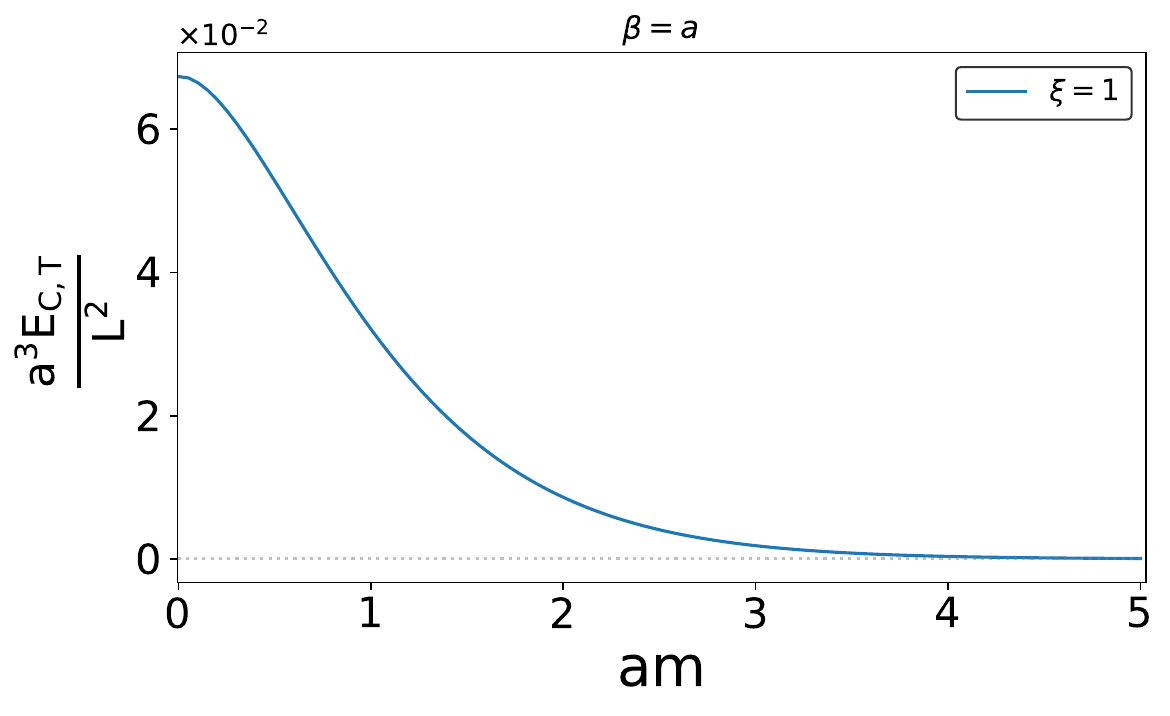}
		\caption{These plots present the behavior of the thermal correction of the Casmir energy as function of dimensionless parameter $ma$, considering $\mu=0$ for the case where the Lorentz symmetry is preserved, $\xi=1$. In the left plot we assumed $\beta=10a$, being $a$ the distance between the two parallel flat planes, and in the right plot, we assumed $\beta=a$.}
		\label{fig1}
	\end{center}
\end{figure}

Unfortunately we could not find in literature a closed expression for the double integral present in \eqref{Cas_5} for arbitrary value of $\xi$.  In this way we understand that this expressions is not very enlightening. In order to provide a better understanding  of thermal correction for the Casimir energy, $E_{C,T}$, in Fig. \ref{fig1} we exhibit the behavior of $a^3E_{T,C}/L^2$  as function of $ma$ considering $\xi=1$ for $\beta=10a$, low temperature, and $\beta=a$, high temperature, assuming $\mu=0$.

\begin{figure}[!htb]
	\begin{center}
		\includegraphics[scale=0.35]{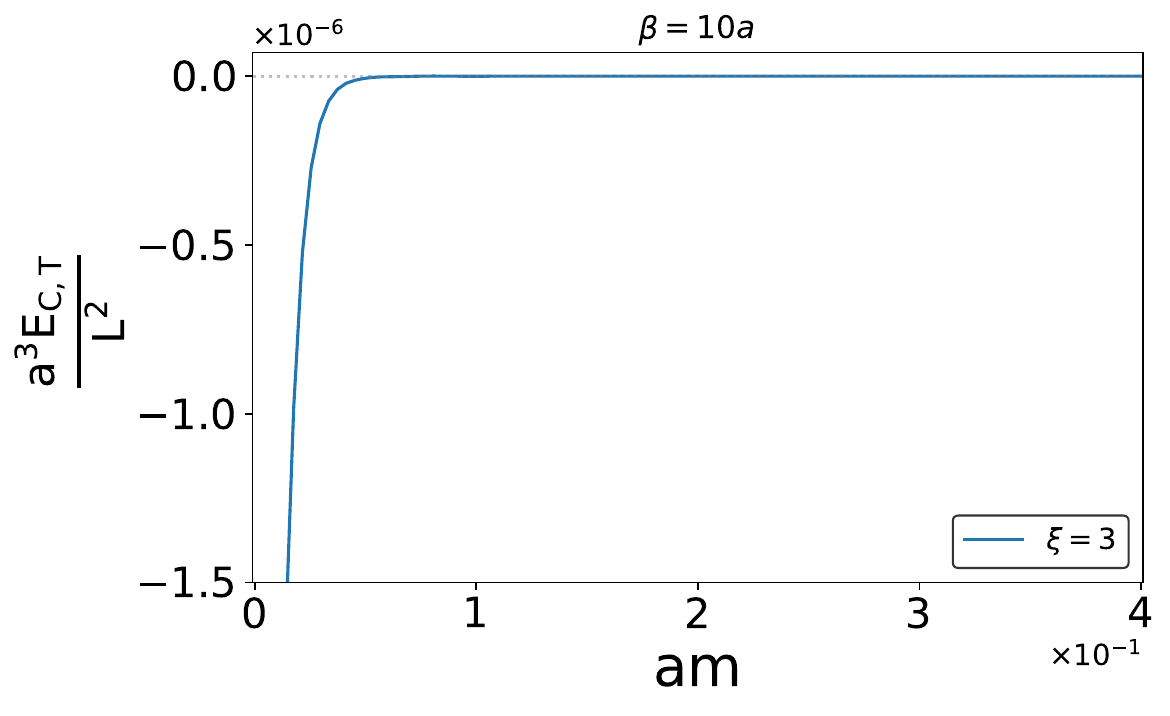}
		\quad
		\includegraphics[scale=0.35]{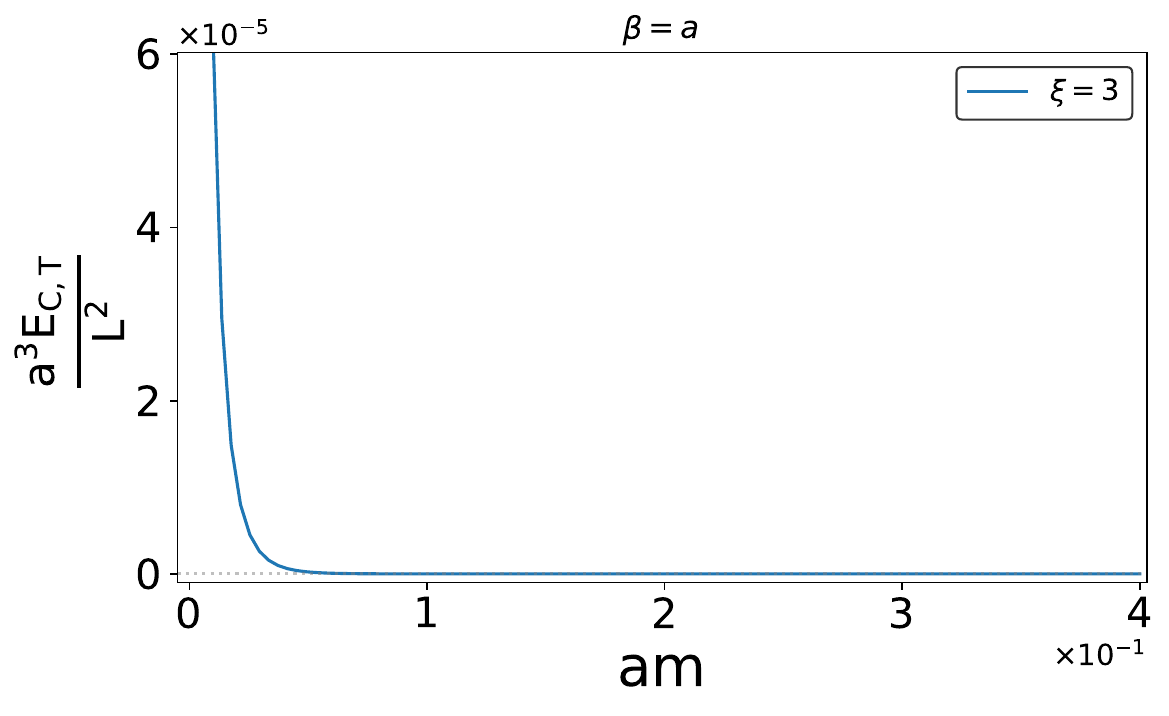}
		\vfill
		\includegraphics[scale=0.35]{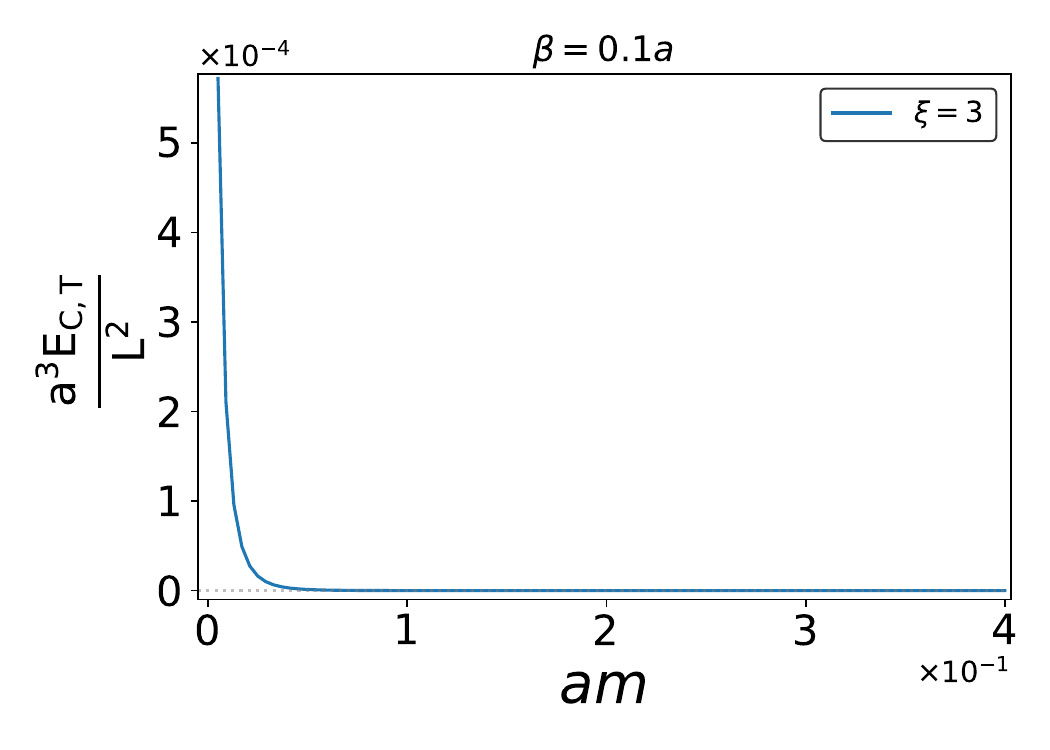}
		\caption{These plots present the behavior of the thermal correction of the Casmir energy as function of $ma$ for the case of LV scenario, $\xi=3$, considering $a/l=10^2$. In the left plot on the top we assume $\beta=10a$ and in the right one we assumed $\beta=a$. For the plot on the bottom we take $\beta=a/10$. For all plots we assumed $\mu=0$.}
		\label{fig2}
	\end{center}
\end{figure}

In Fig. \ref{fig2}, we present the behavior of \eqref{Cas_5} for $\xi=3$, assuming three distinct values for $\beta$: $\beta=10a$, $\beta=a$ and $\beta=a/10$, taking $\mu=0$. In these plots, we assume as an illustrative example that $a/l=10^2$. Two aspects deserve to be mentioned in this graph: the decrease in the intensity of $E_{C,T}$ compared with the previous plots, $\xi=1$, and the change of energy correction sign. For $\beta=10a$ the thermal correction is negative, while for the other cases it is positive. It is clear from the plots that the absolute values of the vacuum energy increase as the temperature increases. This happens in all boundary conditions cases, and for different values of $\xi$.
	
\begin{figure}[!htb]
	\begin{center}
		\includegraphics[scale=0.35]{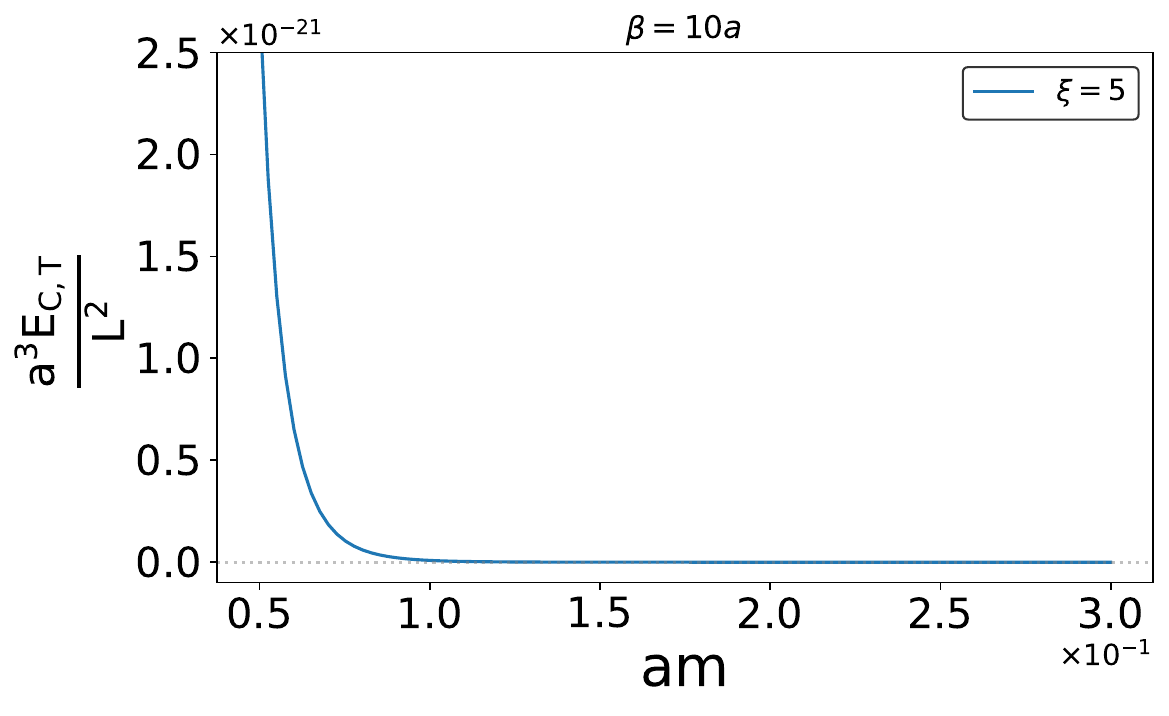}
		\quad
		\includegraphics[scale=0.35]{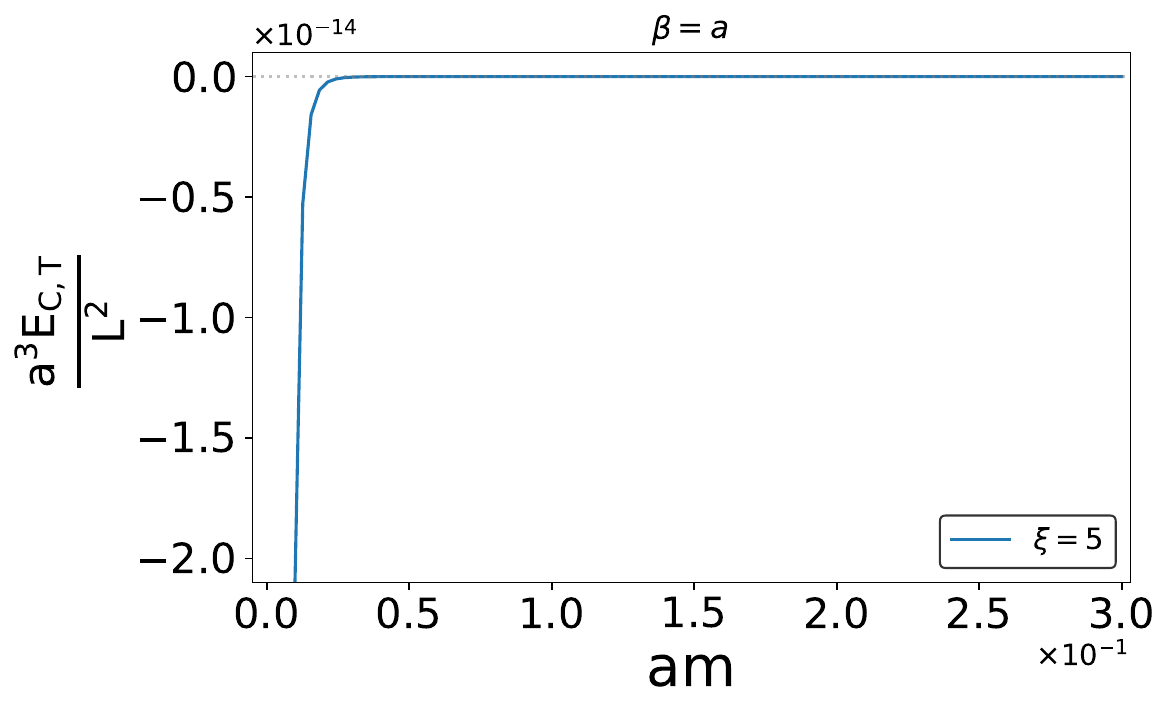}
		\caption{These plots present the behavior of the thermal correction of the Casmir energy as function of $ma$ considering $\xi=5$. In the left plot we assumed $\beta=10a$ and in the right plot, we assumed $\beta=a$. For both plots we assumed $\mu=0$.}
		\label{fig3}
	\end{center}	
\end{figure}

To complete this analysis we exhibit in Fig. \ref{fig3} the behavior of the thermal correction for $\xi=5$, considering $a/l=10^2$. Here we assume only two distinct values for $\beta$: $\beta=10a$ in the left plot and $\beta=a$, in the right one. For both plots we assumed $\mu=0$. The same aspects mentioned in previous plots, also are applicable here.
	
\subsection{Neumann boundary condition}	
\label{Neumann}
In this subsection we want to analyze the thermal correction to the Casimir energy, admitting that the scalar field obeys the Neumann boundary condition on the plates, given by expressions below,
\begin{eqnarray}
	\label{Neumann_condition}
	\begin{aligned}
		\frac{\partial \phi(x)}{\partial z} \Bigg{|}_{z=0} = \frac{\partial \phi(x)}{\partial z} \Bigg{|}_{z=a} = 0.
	\end{aligned}
\end{eqnarray}

The solution of the modified Klein-Gordon equation, Eq. \eqref{kgmodific}, compatible with the above boundary condition is,
\begin{eqnarray}
	\label{Neumann_operator_field}
	\begin{aligned}
		\hat{\phi}(x) = \int d^2\vec{k} \sum_{n=0}^{\infty} c_n \cos \Big{(} \frac{n\pi}{a} z \Big{)} \Big{[} \hat{a}_n(\vec{k})
		e^{-ikx} + \hat{a}_n^{\dagger}(\vec{k}) e^{ikx} \Big{]},
	\end{aligned}
\end{eqnarray}
with the normalization constant given by,
\begin{eqnarray}
	\label{normalization_constant}
	c_{n}=\left \{\begin{array}{c}
	\frac{1}{\sqrt{2(2\pi)^2 a k_0}} \ \ \text{for} \ n=0 \ , \\
		\frac{1}{\sqrt{(2\pi)^2 a k_0}} \ \ \text{for} \text{ } n\geq 0 \  ,
	\end{array} \right.
\end{eqnarray}
	where
\begin{eqnarray}
	\label{k_def}
	kx=k_0x_0-k_{x}x-k_{y}y \  .
\end{eqnarray}

For this case the dispersion relations coincides with the corresponding one given in previous analysis:
\begin{eqnarray}
	\label{k0_def_1}
	k_0=\omega_{{\vec{k}},n}=l^{\xi-1}\sqrt{(k_x^2+k_y^2+(n\pi/a)^2)^\xi+v^{2\xi}} \  ,  
\end{eqnarray}
where ${\cal}{v}^\xi=ml^{1-\xi}$.

The thermal energy reads,
\begin{eqnarray}
	\label{energyT_0}
	E_T=\frac{L^2}{(2\pi)^2}\sum_{j=1}^{\infty}e^{j\beta \mu}\int_{0}^{\infty}{d^2k}\sum_{n=0}^{\infty}\omega_{\vec{k},n}e^{-j\beta \omega_{\vec{k},n}} \   .
\end{eqnarray}
So in order to develop the summation over the quantum number $n$ we will use the Abel-Plana summation formula below,
\begin{eqnarray}
	\label{APN}
	\sum_{n=0}^{\infty}F(n)=\frac{1}{2}F(0)+\int_{0}^{\infty}{dt}F(t)+i\int_{0}^{\infty}\frac{dt}{e^{2\pi t}-1}\left[F(it)-
	F(-it)\right]  \  .
\end{eqnarray}
Substituting \eqref{APN} into \eqref{energyT_0}, three distinct terms take places. Comparing these contributions with the corresponding one for Dirichlet boundary condition case, we see that the only difference is in the sign of the first contribution, that for this case it is positive, resulting in 
	\begin{eqnarray}
	\label{Ener_1_M}
	E_T^{(1)}&=&\frac{L^2l^{\xi-1}}{4\pi\xi}\frac{2^{1/\xi}\Gamma(1/\xi)}{\sqrt{2\pi}}\sum_{j=1}^\infty e^{j\beta\mu}\left[v^{1+3\xi/2}\frac{K_{1/\xi+3/2}(j\beta m)}{(j\beta l^{\xi-1})^{1/\xi-1/2}}\right.\nonumber\\
	&-&\left.v^{1+\xi/2}\frac{K_{1/\xi+1/2}(j\beta m)}{(j\beta l^{\xi-1})^{1/\xi+1/2}}\right]  \  .
\end{eqnarray}
The other terms remain the sames. For that reason,  we will not repeat all the calculations.

\subsection{Mixed boundary condition}	
\label{Mixxed}
Now, let us consider a scalar field that obeys a Dirichlet boundary condition on one plate and a Neumann boundary condition 
on the other one. For this case, two different configurations take place:
	\begin{eqnarray}
		\phi(\vec{x}) \Big{|}_{z=0} = \frac{\partial \phi(\vec{x})}{\partial z} \Bigg{|}_{z=a} = 0.
	\end{eqnarray}
	\begin{eqnarray}
		\frac{\partial \phi(\vec{x})}{\partial z} \Bigg{|}_{z=0} = \phi(\vec{x}) \Big{|}_{z=a} = 0.
	\end{eqnarray}

The solution of the modified Klein-Gordon equation considering the above conditions, are:
\begin{eqnarray}
	\label{Misto_campo_operator_a}
	\hat{\phi}_a(x) = \int d^2\vec{k} \sum_{n=0}^{\infty} \frac{1}{\sqrt{(2\pi)^2ak_0}} \sin \Big{[} \Big{(}
	n+\frac{1}{2} \Big{)}\frac{\pi}{a}z \Big{]} \Big{[} \hat{a}_n(\vec{k})e^{-ikx} + \hat{a}_n^{\dagger}(\vec{k})e^{ikx} \Big{]}  \  ,
\end{eqnarray}
for the first configuration and
\begin{eqnarray}
	\label{Misto_campo_operator_b}
	\hat{\phi}_b(x) = \int d^2\vec{k} \sum_{n=0}^{\infty} \frac{1}{\sqrt{(2\pi)^2ak_0}} \cos \Big{[} \Big{(}
	n+\frac{1}{2} \Big{)}\frac{\pi}{a}z \Big{]} \Big{[} \hat{a}_n(\vec{k})e^{-ikx} + \hat{a}_n^{\dagger}(\vec{k})e^{ikx} \Big{]} \  ,
\end{eqnarray}
for the second configuration. Nonetheless both field operators, $\hat{\phi}_a(x)$ and $\hat{\phi}_a(x)$, provide the same dispersion relation, 
\begin{eqnarray}
	\label{Rel_Mix}
	k_0=\omega_{{\vec{k}},n}=l^{\xi-1}\sqrt{(k_x^2+k_y^2+((n+1/2)\pi/a)^2)^\xi+v^{2\xi}} \  .
\end{eqnarray}

The  thermal  energy associated with this system reads,
\begin{eqnarray}
	\label{energyT0}
	E_T=\frac{L^2}{(2\pi)^2}\sum_{j=1}^{\infty}e^{j\beta \mu}\int_{0}^{\infty}{d^2k}\sum_{n=0}^{\infty}\omega_{\vec{k},n}e^{-j\beta \omega_{\vec{k},n}} \   .
\end{eqnarray}

Now in order to develop the summation over the quantum number $n$ we use the Abel-Plana summation formula for half-integer numbers \cite{{Bordag:2009zzd,Saharian:2007ph}}:
\begin{eqnarray}
	\label{Abel_Plana_2}
	\begin{aligned}
		\sum_{n=0}^{\infty} F(n+1/2) = \int_0^{\infty} F(t)dt - i\int_0^{\infty} \frac{dt}{e^{2\pi t}+1}
		\Big{[} F(it) - F(-it) \Big{]} \  .
	\end{aligned}
\end{eqnarray}

Substituting \eqref{Rel_Mix} into \eqref{energyT0}, and using the Abel-Plana summation formula above, we see that the thermal energy presents two distinct contributions. We can see that the first one provides the same expression given in \eqref{Ener_2} for the energy density. So we will not repeat the whole calculation. It remains only to obtain the contribution given by the second term of the Abel-Plana formula above.

\subsubsection{The thermal Casimir energy correction}
\label{Casimir_Mixed}
Here we will analyze the thermal correction to the Casimir energy for the system under investigation. As we have already mentioned this term comes from the substitution of the second term of \eqref{Abel_Plana_2} into \eqref{energyT0}. It reads,
	\begin{eqnarray}
	\label{Cas_M}
	E_{T,C}&=&-\frac{iL^2l^{\xi-1}}{2\pi^2}\frac a\pi\sum_{j=1}^\infty e^{j\beta\mu}\int d^2{\vec{k}}\int_0^\infty \frac {du}{e^{2au}+1} \left[\sqrt{\left(k^2+(iu)^2\right)^\xi+v^{2\xi}}e^{-j\beta l^{\xi-1}\sqrt{\left(k^2+(iu)^2\right)^\xi+v^{2\xi}}}\right.\nonumber\\
	&-&\left.\sqrt{\left(k^2+(-iu)^2\right)^\xi+v^{2\xi}}e^{-j\beta l^{\xi-1}\sqrt{\left(k^2+(-iu)^2\right)^\xi+v^{2\xi}}}\right]  \  ,
\end{eqnarray}
where we have introduced a new variable $u=\pi t/a$.

Following the same steps as in the development of \eqref{Cas_5} we divide the integral with respect to the variable $u$ into two sub-intervals.  The integral over the first segment, i.e., from $[0, \sqrt{k^2+v^2}]$ vanishes, remaining only the integral over the second segment, from $[\sqrt{k^2+v^2}, \infty)$. In addition we can integrate  over the angular variable on the plane $(k,\varphi)$. Finally we obtain,
\begin{eqnarray}
	\label{Cas_2_M}
	E_{T,C}&=&\frac{aL^2l^{\xi-1}}{\pi^2}\sin(\pi\xi/2)\sum_{j=1}^\infty e^{j\beta\mu}\int_0^\infty dk k\int_{\sqrt{k^2+v^2}}^\infty \frac {du}{e^{2au}+1} \sqrt{\left(u^2-k^2\right)^\xi-v^{2\xi}}\nonumber\\
	&\times&\cos\left(j\beta l^{\xi-1}\sqrt{\left(u^2-k^2\right)^\xi-v^{2\xi}}\right)  \  .
\end{eqnarray}

	Defining a new variable $\rho=\sqrt{k^2+v^2}$, we see that $kdk=\rho d\rho$. Our next step is to introduce  a new variable, $u=\rho y$, what leads us to
\begin{eqnarray}
	\label{Cas_4_M}
	E_{T,C}&=&\frac{aL^2l^{\xi-1}}{\pi^2}\sin(\pi\xi/2)\sum_{j=1}^\infty e^{j\beta\mu}\int_v^\infty d\rho \rho^2\int_1^\infty \frac {dy}{e^{2a\rho y}+1} \sqrt{[\rho^2(y^2-1)+v^2]^\xi-v^{2\xi}}\nonumber\\
	&\times&\cos\left(j\beta l^{\xi-1}\sqrt{[\rho^2(y^2-1)+v^2]^\xi-v^{2\xi}}\right)  \  .
\end{eqnarray}
Finally defining a new variable $\rho=vz$, we have:
\begin{eqnarray}
	\label{Cas_5_M}
	E_{T,C}&=&\frac{L^2(am)^{1+3/\xi}\rho_\xi^3}{a^3\pi^2}\sin(\pi\xi/2)\sum_{j=1}^\infty e^{j\beta\mu}\int_1^\infty dz z^2\int_1^\infty \frac {dy}{e^{2(ma)^{1/\xi}\rho_\xi z y}+1} \nonumber\\
	&\times&\sqrt{[z^2(y^2-1)+1]^\xi-1}\cos\left(j\beta m\sqrt{[z^2(y^2-1)+1]^\xi-1}\right)  \ ,
\end{eqnarray}
being $\rho_\xi=(a/l)^{1-1/\xi}$.\footnote{At this point the sames comments mentioned after \eqref{Cas_5}, could be included; however we understand be totally unnecessary.} 

The Casimir energy thermal correction can be presented in a closed formula fo the case $\xi=1$. In Appendix \ref{Ap_b}, we briefly present the steps necessary to obtain this expression. Our result is:
\begin{eqnarray}
	\label{Ener-Closed_M}
	E_{C,T}&=&-\frac{L^2am^4}{\pi^2}\sum_{j=1}^\infty e^{j\beta \mu}\frac{\partial}{\partial\alpha}\left[\alpha\sum_{n=1}^\infty \frac{(-1)^n K_2\left(\sqrt{4n^2(ma)^2+\alpha^2}\right)}{(4n^2(ma)^2+\alpha^2)}\right]|_{\alpha=j\beta m} \  . 
\end{eqnarray}

\begin{figure}[!htb]
	\begin{center}
		\includegraphics[scale=0.35]{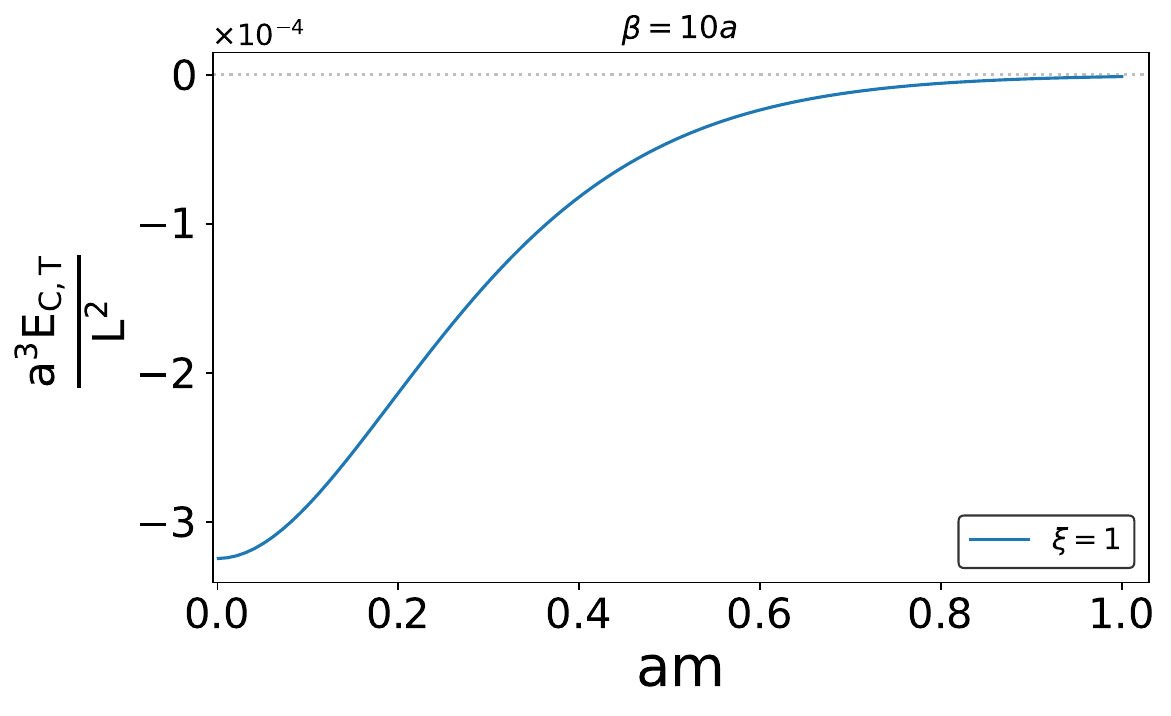}
		\quad
		\includegraphics[scale=0.35]{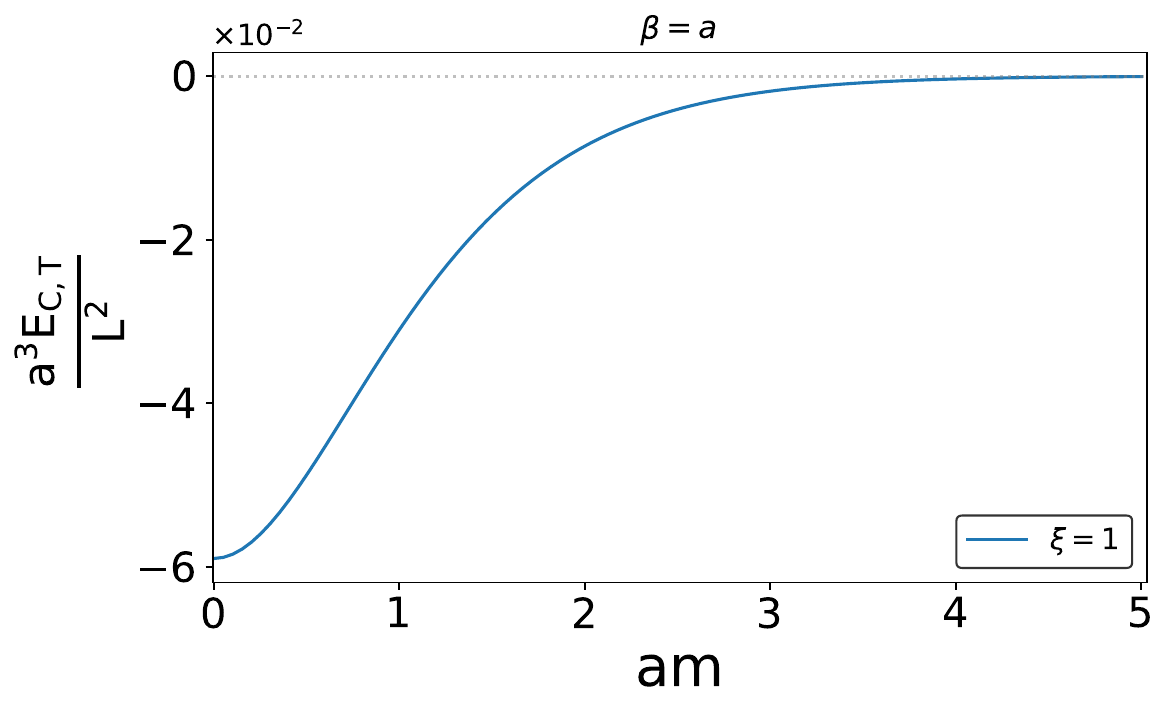}
		\caption{These plots exhibit behavior of \eqref{Cas_5_M} as function of $ma$ for the case $\xi=1$. In the left panel we assumed $\beta=10a$, and in the right panel we assumed $\beta=a$.}
		\label{fig4}
	\end{center}
\end{figure}

Unfortunately we also could not find in literature a closed expression for the double integral presented in \eqref{Cas_5_M}	for arbitrary value of $\xi$. So, as in the analysis of behavior of the thermal Casimir energy for Dirichlet fields, here  we develop some numerical evaluation exhibiting the behavior of  $a^3E_{T,C}/L^2$  as function of $ma$ for three different values of the critical exponent, $\xi=1, \ 3 \ {\rm and} \ 5$, and for $\beta=10a$ and $\beta=a$.  In all the following plots we assume $\mu=0$.  In Fig. \ref{fig4} we present two plots considering $\xi=1$.

\begin{figure}[!htb]
	\begin{center}
		\includegraphics[scale=0.35]{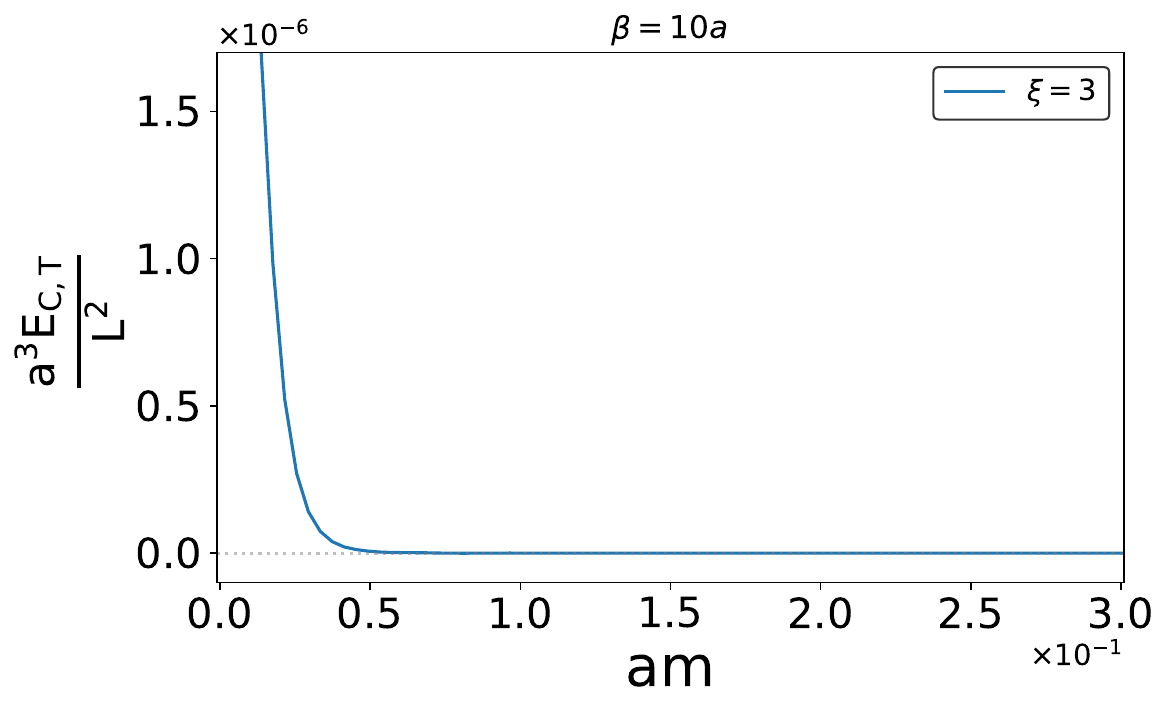}
		\quad
		\includegraphics[scale=0.35]{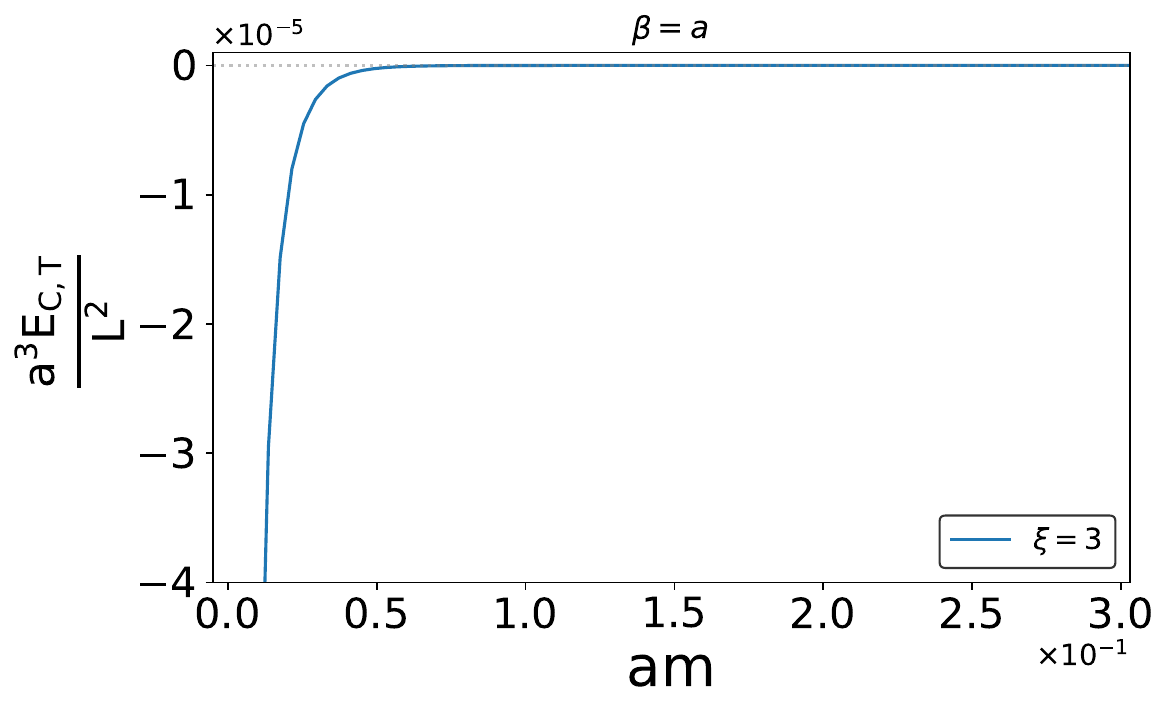}
		\caption{These plots represent the behavior of the thermal correction of the Casmir energy for mixed boundary condition as function of $ma$ for $\xi=3$. In the left panel we assume $\beta=10a$, and in the right one, we assume $\beta=a$.}
		\label{fig5}
	\end{center}
\end{figure}

\begin{figure}[!htb]
	\begin{center}		\includegraphics[scale=0.35]{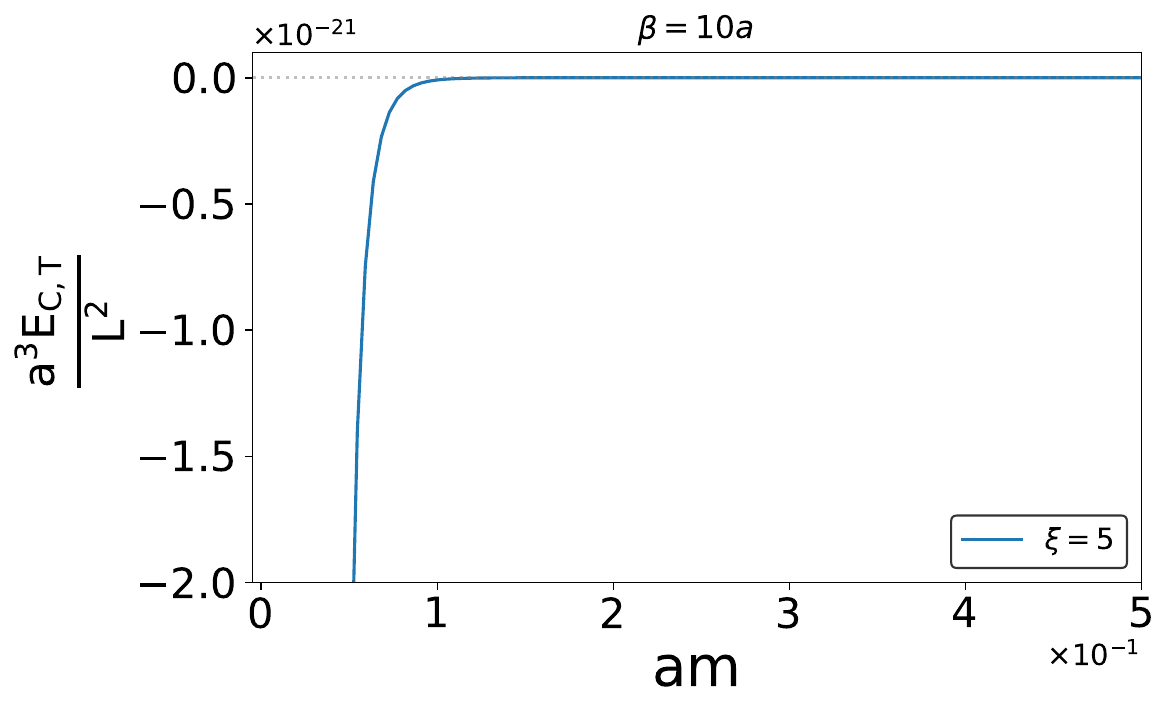}
		\quad
		\includegraphics[scale=0.35]{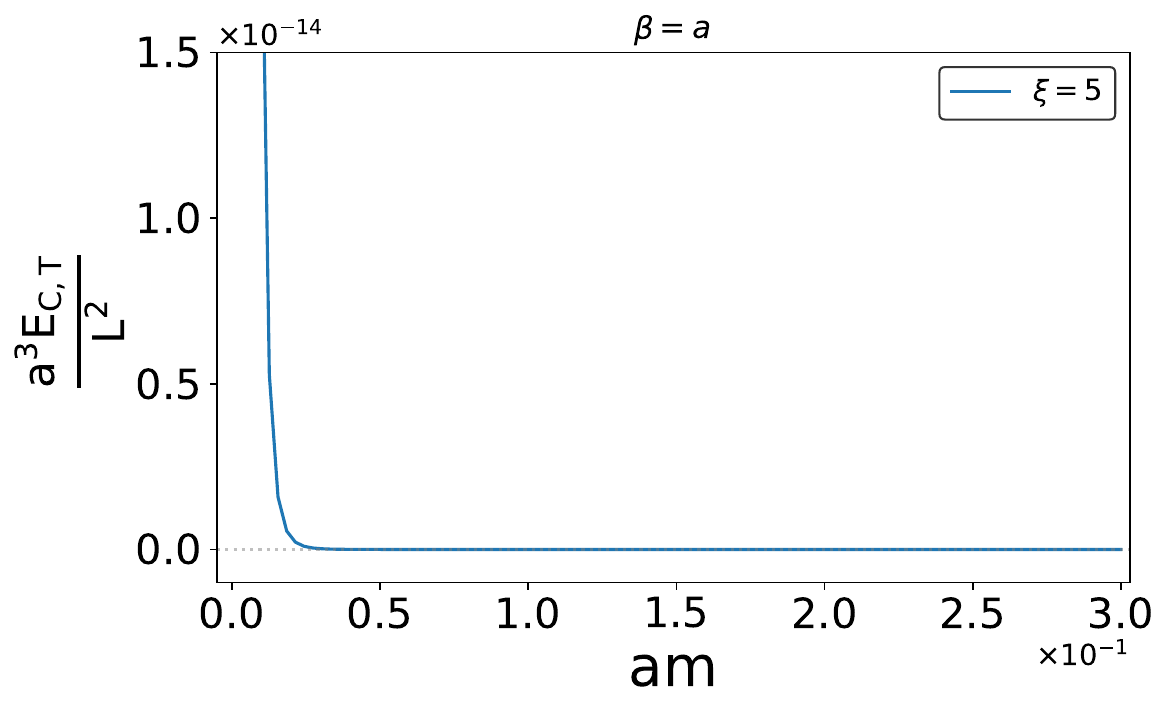}
		\caption{These plots present the behavior of the thermal correction of the Casmir energy as function of $ma$ for $\xi=5$. In the left plot we assumed $\beta=10a$ and in the right one, we assumed $\beta=a$.}
		\label{fig6a}
	\end{center}
\end{figure}

In Fig. \ref{fig5}, we exhibit  the behavior of \eqref{Cas_5_M} as function of $am$ for $\xi=3$, assuming $\beta=10a$ and $\beta=a$. Finally in Fig. \ref{fig6a}, we also exhibit  the behavior of the thermal correction for $\xi=5$ for the same temperatures. In both figures we adopted $a/l=10^2$. 

\section{Conclusion}
\label{Concl}
In this paper we have analyzed the effect of the temperature on the Casimir energy associated with real quantum scalar field, in a Ho{\v{r}}ava-Lifshitz Lorentz violation symmetry scenario. In order to do that we assumed that the field is confined in the region between two large and parallel plates, by imposing that the field obeys Dirichlet, Neumann and mixed boundary conditions on them. In order to develop these calculations we had to obtain the corresponding dispersion relations. Differently the case of a Lorentz preserved scenario, the dispersion relations depend on the parameter $l$, and also on the parameter $\xi\neq 1$, named critical exponent,  responsible for the Lorentz violation. The boundary conditions imposed restrictions on the momentum perpendicular to the plates, and to develop the summation over these discrete quantum number, we applied the Abel-Plana summation formula. In this way, we were able to present the thermal energy densities and mainly the thermal Casimir energies. As to the thermal energy density we provided closed expressions given in terms of Macdonald functions; as to the Casimir energy, only not solvable integral representations were possible to be obtained.

Because the analytic expressions for  thermal Casimir energies corrections  are not very enlightening, we decided to  furnish numerical plots that exhibit the behavior of them as function of the product $ma$, considering different values of temperature and critical exponent. From these graphs we observed that the intensity of the thermal Casimir energy decreases by many orders of magnitude when the critical exponent increases. In addition, for a fixed value of $\xi$ the intensities of the thermal correction increases with the temperature as expected.

In general the analysis of Casimir effects results from energy of the vacuum state of a quantum field in the presence of boundaries. All the excited states are neglected. In practice, the appropriated state of a quantum system contain real particles, this is typical situation of state containing particle in thermal equilibrium. In this way,  ones has to consider an ensemble of state characterized by a temperature $T=\frac1\beta$. Another simplified situation takes place considering QFT in a Lorentz invariant symmetry scenario; however, accepting that the Lorentz symmetry is violated as proposed in the HL  model, the more relevant contribution would be for $\xi=3$. Finally both contributions, finite temperature and Lorentz violation, can considered as part of the source in the $1\%$  of the experimental error estimated in \cite{Harris}.  In this sense our work comes to add additional explanation for the experimental error present in the measurement of Casimir energy

\section*{Acknowledgments}
 H.M. is partially supported by the Brazilian agency National Council for Scientific and Technological Development (CNPq) under Grant No. 308049/2023-3.
\appendix
\section{Development of double integrals for Dirichlet case for $\xi=1$}
\label{Ap_a}
The double  integral in \eqref{Cas_5}, for the case $\xi=1$, can be expressed by,
\begin{eqnarray}
	\label{Int_1B}
	{\cal{I}}_1&=&\frac{\partial}{\partial \alpha_j}\int_1^\infty dz z^2\int_1^\infty \frac {dy}{e^{2ma z y}-1} \sin(\alpha_jz\sqrt{y^2-1})|_{\alpha_j=j\beta m} \  . 	
\end{eqnarray} 
Using the identity \eqref{identidade}, and defining $y^2=x^2+1$, we obtain the expression,
\begin{eqnarray}
	\label{Int_1C}
	{\cal{I}}_1=\frac{\partial}{\partial \alpha_j}\sum_{n=1}^\infty\int_1^\infty dz z^2\int_0^\infty \frac {dxx}{\sqrt{x^2+1}}{e^{-2nma z \sqrt{x^2+1}}} \sin(\alpha_jxz)|_{\alpha_j=j\beta m} \  . 	
\end{eqnarray} 
Using the integral representation for the Macdonald function \cite{Grad},
\begin{eqnarray}
	\label{Int_rep}
	K_1\left(\gamma\sqrt{\beta^2+b^2}\right)=\frac{\sqrt{\beta^2+b^2}}{\gamma b}\int_0^\infty dx x \frac{e^{-\beta\sqrt{\gamma^2+x^2}}}{\sqrt{\gamma^2+x^2}}\sin(bx)  \  ,
\end{eqnarray}
the expression \eqref{Int_1C} can be written as,
\begin{eqnarray}
	\label{Int_1D}
	{\cal{I}}_1=\frac{\partial}{\partial \alpha_j}\sum_{n=1}^\infty \frac{\alpha_j}{\sqrt{4n^2(am)^2+\alpha_j^2}}\int_1^\infty dz z^2K_1\left(z\sqrt{4n^2(am)^2 +\alpha_j^2}\right)|_{\alpha_j=j\beta m} \  . 	
\end{eqnarray} 
Defining a new variable $u=z\sqrt{4n^2(am)^2 +\alpha_j^2}$, we have,
\begin{eqnarray}
	\label{Int_1E}
	{\cal{I}}_1=\frac{\partial}{\partial \alpha_j}\sum_{n=1}^\infty \frac{\alpha_j}{(4n^2(am)^2+\alpha_j^2)^2}\int_{\sqrt{4n^2(am)^2 +\alpha_j^2}}^\infty du u^2K_1\left(u\right)|_{\alpha_j=j\beta m} \  . 	
\end{eqnarray} 
Using the identity,
\begin{eqnarray}
	\frac{\partial}{\partial u}(u^\nu K_\nu(u))=-u^\nu K_{\nu-1}(u)  \  , 
\end{eqnarray}
we can integrate over the new variable $u$, resulting in,
\begin{eqnarray}
	\label{Int_1F}
	{\cal{I}}_1=\frac{\partial}{\partial \alpha_j}\alpha_j\sum_{n=1}^\infty \frac{K_2\left(z\sqrt{4n^2(am)^2 +\alpha_j^2}\right)}{4n^2(am)^2+\alpha_j^2}|_{\alpha_j=j\beta m} \  . 	
\end{eqnarray} 

\section{Development of double integrals for mixed case for $\xi=1$}
\label{Ap_b}
The double  integral in \eqref{Cas_5_M}, for the case $\xi=1$, can be expressed by,
\begin{eqnarray}
	\label{Int_1B_M}
	{\cal{I}}_2&=&\frac{\partial}{\partial \alpha_j}\int_1^\infty dz z^2\int_1^\infty \frac {dy}{e^{2ma z y}+1} \sin(\alpha_jz\sqrt{y^2-1})|_{\alpha_j=j\beta m} \  . 	
\end{eqnarray} 
Using the relation
\begin{eqnarray}
	\frac1{e^x+1}=-\sum_{n=1}^\infty (-1)^n e^{-nx} \  ,
\end{eqnarray}
and defining $y^2=x^2+1$, the expression \eqref{Int_1B_M} becomes, 
\begin{eqnarray}
	\label{Int_1C_M}
	{\cal{I}}_2=-\frac{\partial}{\partial \alpha_j}\sum_{n=1}^\infty(-1)^n\int_1^\infty dz z^2\int_0^\infty \frac {dxx}{\sqrt{x^2+1}}{e^{-2nma z \sqrt{x^2+1}}} \sin(\alpha_jxz)|_{\alpha_j=j\beta m} \  . 	
\end{eqnarray} 	
Again, using the integral representation for the Macdonald function, \eqref{Int_rep}, and following all the steps presented in Appendix \ref{Ap_a}, we arrive at,
\begin{eqnarray}
	\label{Int_1F_M}
	{\cal{I}}_2=-\frac{\partial}{\partial \alpha_j}\alpha_j\sum_{n=1}^\infty \frac{(-1)^n K_2\left(z\sqrt{4n^2(am)^2 +\alpha_j^2}\right)}{4n^2(am)^2+\alpha_j^2}|_{\alpha_j=j\beta m} \  . 	
\end{eqnarray}

\end{document}